\documentclass[pra,amsmath,amssymb,amsfonts,superscriptaddress,eqsecnum,twocolumn]{revtex4-2}
\usepackage{graphicx,color,mathtools,bm,braket,amsmath,newtxtext,newtxmath}

\usepackage[hyphenbreaks]{breakurl}
\usepackage[colorlinks=true,linkcolor=blue,citecolor=blue,urlcolor =blue]{hyperref}

\newcommand{\op}[1]{\hat{#1}}

\begin{document}

\title{Quantum squeezing via self-induced transparency in a photonic crystal fiber}

\author{M. S. Najafabadi}  
\email{mojdeh.shikhali-najafabadi@mpl.mpg.de}
\affiliation{Max-Planck-Institut f\"{u}r die Physik des Lichts, 91058~Erlangen, Germany}       

\author{L. L. S\'anchez-Soto}
\affiliation{Max-Planck-Institut f\"{u}r die Physik des Lichts, 91058~Erlangen, Germany}   
\affiliation{Departamento de \'Optica, Facultad de F\'{\i}sica, Universidad Complutense, 28040~Madrid, Spain} 

\author{J. F. Corney}
\affiliation{School of Mathematics and Physics, University of Queensland, Brisbane, Queensland 4072, Australia} 

\author{N. Kalinin}
\affiliation{Max-Planck-Institut f\"{u}r die Physik des Lichts, 91058~Erlangen, Germany}    
\affiliation{Institut f\"{u}r Optik, Information und Photonik,  Friedrich-Alexander-Universit\"{a}t Erlangen-N\"{u}rnberg, 91058~Erlangen, Germany}

\author{A. A. Sorokin}
\affiliation{A.V. Gaponov-Grekhov Institute of Applied Physics of the Russian Academy of Sciences,  46 Ulyanov Str, 603950 Nizhny Novgorod, Russia}
        
\author{G. Leuchs}      
\affiliation{Max-Planck-Institut f\"{u}r die Physik des Lichts, 91058~Erlangen, Germany}        
\affiliation{Institut f\"{u}r Optik, Information und Photonik,  Friedrich-Alexander-Universit\"{a}t Erlangen-N\"{u}rnberg, 91058~Erlangen, Germany}         

\begin{abstract}
We study the quantum squeezing produced in self-induced transparency in a photonic crystal fiber by performing a fully quantum simulation based on the positive $P$ representation. The amplitude squeezing  depends on the area of the initial pulse: when the area is $2\pi$, there is no energy absorption and no amplitude squeezing. However, when the area is between 2$\pi$ and 3$\pi$, one observes amplitude-dependent energy absorption and a significant amount of squeezing. We also investigate the effect of damping and temperature: the results indicate that a heightened atom-pulse coupling, caused by an increase in the spontaneous emission ratio reduces the  amplitude squeezing.
\end{abstract}

\date{\today} \maketitle 

\section{Introduction}

Self-induced transparency (SIT) in two-level atomic systems is one of the most well-known coherent pulse propagation phenomena: above a certain intensity threshold, the absorption of a pulse by resonant transitions decreases strongly and the medium becomes almost completely transparent, which is accompanied by a considerable reduction in the group velocity (for reviews see~\cite{Lamb:1971aa,Slusher:1974aa,Poluektov:1975aa,Maimistov:1990aa}).  This was first reported by McCall and Hahn~\cite{McCall:1967aa,McCall:1969aa}, who, using a semiclassical description, demonstrated that the two-level medium becomes transparent to a 2$\pi$ pulse through a strong absorption. This semiclassical model is nowadays standard in quantum-optics textbooks that study the effects of atomic coherence~\cite{Allen:1975aa,Mandel:1995aa,Scully:1997aa}.

The SIT solitons have been proposed as candidates for pulsed squeezed state generation~\cite{Watanabe:1989aa}, quantum nondemolition measurements~\cite{Drummond:1993aa}, and quantum information storage and retrieval~\cite{Bullough:2004aa}. Moreover, with the recent advances in microstructured fiber technologies~\cite{Russell:2006va}, the generation of squeezing via SIT solitons inside gas-filled single-mode photonic crystal fibers is being considered~\cite{Zhong:2007aa}, which simplifies transverse effects.

In all these advances, the quantum noise and the quantum correlations play a dominant role that cannot be captured by any semiclassical approach. Therefore, a full quantum approach to SIT is an essential step toward a complete understanding of the physics involved.  A  theory of SIT using a linearization \emph{ansatz} has been developed~\cite{Lai:1990aa}  within the framework of the inverse-scattering method~\cite{Shabat:1972aa}.  A further refinement was suggested by using a   coarse-grain-averaged light-atom interaction~\cite{Lee:2009aa} and treating the quantum noise by the back-propagation method~\cite{Lai:1995aa}, which can take into account the field continuum contributions and the atomic fluctuations generally.

These results, important as they are, do not provide proper guidelines for realistic experiments, because they fail to account for any limitations on the squeezing. In this paper, we take an alternative route and adapt a method to deal with the propagation of radiation in an optically pumped two-level medium that has collisional and radiative damping~\cite{Drummond:1991aa}. The idea is to derive a set of stochastic $c$-number differential equations that are equivalent to the Heisenberg operator equations. This is accomplished through use of the positive $P$ representation~\cite{Drummond:1980aa}, which provides a probabilistic description in which stochastic averages corresponds to normally ordered correlations. The method has the advantage of yielding  equations that may be solved numerically, while keeping the key elements that characterize the nonclassical nature of the field.

On the experimental side, when sending a light pulse through an atomic ensemble, the response of each atom will depend on the field amplitude at its respective position, leading to a transverse structure in the resulting light field and in the atomic ensemble as well. This situation changes, if the atoms interact with only one single mode of the field. Then, each photon interacts with all atoms and no transverse structure will develop. To achieve this situation, one might think, e.g., of a glass capillary as a wave guide, but this would not be a good waveguide as it is lossy by coupling to modes propagating to the sides out of the capillary. Lossless guiding by total internal reflection requires a higher index in the core, which is not possible with a simple capillary. But a photonic crystal fibre (PCF) \cite{Russell:2006va} provides both, nearly lossless guiding and a hollow core for the atom vapour. An additional advantage of a PCF is that the decay of the atoms in the core into modes other than the the single longitudinal mode is largely suppressed.

The plan of this paper is as follows. In Sec.~\ref{sec:model} we introduce the model Hamiltonian, investigating how the quantum noise sources arise as coming from both damping and nonlinearities in the Hamiltonian. We explore the dynamics by numerically solving the fully nonlinear stochastic differential equations emerging from the positive $P$ representation. Additional effect of damping, reservoir noise and atom-field coupling is also investigated. In Sec.~\ref{sec:res} we present the main results of our model. We show that the pulse area indeed is the crucial quantity in observing the amplitude squeezing for SIT solitons. Due to the complexity of the dynamics in the expanded phase space, one needs a high number of samples to increase the accuracy of the method.  We discuss the effect of pulse reshaping on the squeezing, as well as the role of damping and temperature on the amplitude squeezing. Our results indicate that the stronger the atom-field coupling (due to an elevated spontaneous emission ratio ), the less the amplitude squeezing. Finally, our concluding remarks are presented in Sec.~\ref{sec:conc}.

\section{Model}
\label{sec:model}

\subsection{Hamiltonian}
\begin{figure}[t]
    \includegraphics[width=\columnwidth]{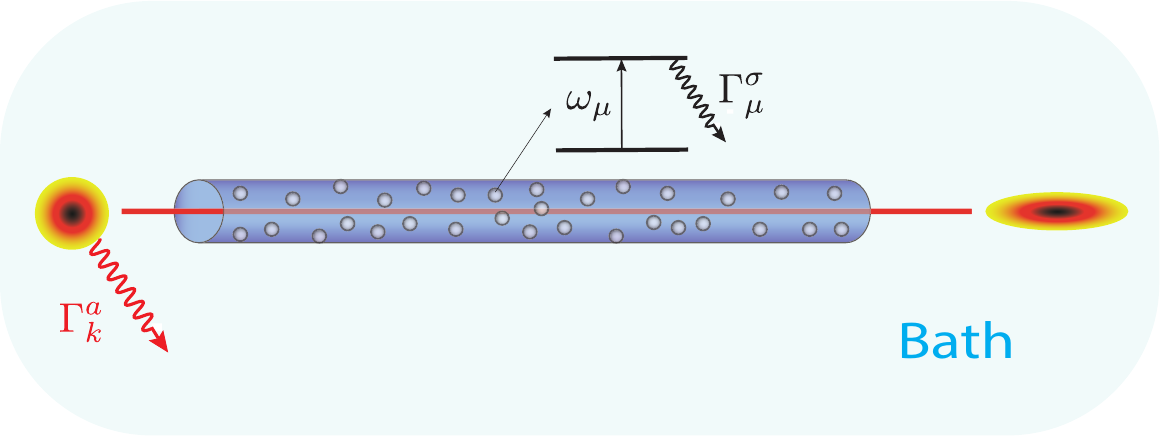}
    \caption{Schematics of the propagation of a coherent pulse in a medium consisting of $N$ two-level atoms in a hollow core fiber. The system (including both the coherent field and the atoms) interacts with the bath,  but the interaction between the atoms is considered negligible.}
    \label{fig:Fig1}
\end{figure}

Following the ideas of Ref.~\cite{Drummond:1991aa}, we first introduce a suitable Hamiltonian which describes the interaction of an ensemble of two-level atoms with a single mode of the radiation field. A schematic picture of our model is shown in Figure.~\ref{fig:Fig1}. In the the rotating-wave and dipole approximations the model Hamiltonian reads as:
\begin{equation}
\label{eq:H}
    \op{H} = \op{H}_{\mathrm{A}}+\op{H}_{\mathrm{F}}+ \op{H}_{\mathrm{B}} + \op{H}_{\mathrm{FB}} + \op{H}_{\mathrm{AB}} + \op{H}_{\mathrm{AF}} 
\end{equation}
where
\begin{equation}
\begin{aligned}
    & \hat{H}_\mathrm{A}  = \frac{1}{2} \sum_{\mu} \hbar \omega_{\mu} \hat{\sigma}^{z}_{\mu},\\
    & \hat{H}_\mathrm{F}  = \sum_{k} \hbar \omega_{k} \hat{a}^{\dag}_{k} \hat{a}_{k},\\
    & \hat{H}_\mathrm{B}  =\hat{H}^{a}+\hat{H}^{\sigma}+ \hat{H}^{z},\\
    & \hat{H}_{AF}  =\hbar \sum_{k} \sum_{\mu} (g\hat{a}^{\dag}_{k}\hat{\sigma}^{-}_{\mu} e^{-ik \cdot z_{\mu}} + \mathrm{H.c.}), \\
    & \hat{H}_\mathrm{AB}  = \hbar \sum_{\mu} (\hat{\Gamma}_{\mu}^{\sigma^{\dag}} \hat{\sigma}_{\mu}^{-} + \hat{\Gamma}_{\mu}^{\sigma} \hat{\sigma}_{\mu}^{+} + \hat{\Gamma}_{\mu}^{z} \hat{\sigma}_{\mu}^{z}),\\
    & \hat{H}_\mathrm{FB}  = \hbar \sum_{k} (\hat{\Gamma}_{k}^{a^{\dag}}\hat{a}_{k} + \hat{\Gamma}_{k}^{a} \hat{a}_{k}^{\dag}).
\end{aligned}
\end{equation}
Here, $\hat{H}_\mathrm{A}$ is the free Hamiltonian of the atoms, with $\omega_{\mu}$ the resonant frequency of the $\mu$th atom described in terms of the standard Pauli operators~\cite{Gardiner:2004aa}, and $\hat{H}_\mathrm{F}$ is the free Hamiltonian of the paraxial field modes propagating in the fiber, each one having frequency $\omega_{k}$ and with annihilation operator $\hat{a}_{k}$ (for a single polarization).

The piece $\hat{H}_\mathrm{B}$ is the free Hamiltonian of the baths corresponding to field modes $\hat{H}^{a}$, atomic dipoles $\hat{H}^{\sigma}$, and collisions $\hat{H}^{z}$. In addition,  $\hat{H}_{\mathrm{AF}}$ is the interaction of the paraxial field with dipole-field coupling $g$; $\hat{H}_{\mathrm{AB}}$ is the interaction of atomic and collisional reservoirs with atoms and, finally, $\hat{H}_{\mathrm{FB}}$ is the interaction of the background reservoir with the radiation field.

Let us briefly discuss the physics behind the Hamiltonian.~\eqref{eq:H}. The paraxial modes are coupled to a background of absorbing dipoles $\hat{\Gamma}_{k}^{a}$, with free Hamiltonian $\hat{H}^{a}$. This describes background absorption and reemission due to other atoms in the medium, as opposed to the resonant ones.

In general, the atoms are also coupled to modes with nonparaxial wave vectors, which form independent radiative reservoirs for each atom, whose operators are $\hat{\Gamma}^{\sigma}_{\mu}$. The free Hamiltonian of these atomic reservoirs is $\hat{H}^{\sigma}$. This approximation of independent reservoirs neglects any transverse dipole-dipole coupling and is thus valid only for relatively low-density optical media, where local-field corrections are negligible. Any optical pumping  is also included in these reservoirs.

Finally, the operators $\hat{\Gamma}_{\mu}^{z}$ describe a coupling of the resonant atoms to a collisional phase-damping reservoir with free Hamiltonian $\hat{H}^{z}$, which describes weak collisions with non-resonant atoms.

To enable a continuous description, we first divide the available volume up into small elements of size $\Delta V$ centered at positions $\mathbf{r}_j$ along the fibre and containing $N_j$ resonant atoms.   The density of resonant atoms in a certain position $\mathbf{r}$ and a certain frequency $\omega$ can then be defined as:
\begin{equation}
    \rho(\mathbf{r}_j, \omega)= \frac{N_j}{\Delta V}f_{\omega}(\omega),
\end{equation}
where $f(\omega)$ is a spectral lineshape~\cite{Hollas:1996aa}. The medium can be considered to be either homogeneously (i.e., Lorentz) or inhomogeneously (i.e., Gaussian) broadened around a central frequency $\omega_{0}$. 

The dipole-field coupling is assumed to be identical for all the atoms and independent of the frequency and the wave vector. For an ideal two-level atom this coupling reads~\cite{Allen:1975aa}
\begin{equation}
     g^2 = \left( \frac{3\gamma_{0}c\lambda_{0}^{2}}{4V}\right) \, ,
\end{equation}  
where $\lambda_{0}$ is the resonant free-space wavelength, $V$ the quantization mode volume, $c$ the speed of light and
\begin{equation}
    \gamma_{0} = \frac{\omega_{0}^{3}}{3\pi\varepsilon_{0} \hbar c^{3}} \mu_{12}^2
\end{equation}
is the spontaneous decay rate (or Einstein $A$ coefficient). Here, $\mu_{12}$ is the relevant dipole matrix element for a linearly polarized pulse.
 
The evolution of the system can be studied by the master equation for the atom-field system by tracing out the reservoir variables and applying the standard Markov approximation~\cite{Gardiner:2004aa}:
\begin{align}
\label{Eq:master_Eq}
  \frac{d \hat{\varrho}}{dt} = \frac{1}{i\hbar} [ \hat{H} , \hat{\varrho} ] + \hat{\mathcal{L}}_{\mathrm{AB}} [ \hat{\varrho} ] + \hat{\mathcal{L}}_{\mathrm{FB}}[ \hat{\varrho}] \, , 
\end{align}
where $\hat{\varrho}$ is the density matrix of the system. The Linbladian superoperators $\hat{\mathcal{L}}_{\mathrm{AB}}$ and $\hat{\mathcal{L}}_{\mathrm{FB}} $ describe relaxation into the reservoir modes in both atomic and field variables and take the form~\cite{Breuer:2002aa}
\begin{widetext}
\begin{equation}
\begin{aligned}
   \hat{\mathcal{L}}_{\mathrm{AB}} [\hat{\varrho}] & = 
    \sum_{\mu}  \tfrac{1}{2} W_{21}   ( [ \hat{\sigma}^{-}_{\mu} \hat{\varrho}, \hat{\sigma}^{+}_{\mu} ] + [ \hat{\sigma}_{\mu}^{-}, \hat{\varrho}\sigma_{\mu}^{+} ] ) 
    + \tfrac{1}{2}W_{12} ( [ \sigma_{\mu}^{+}\hat{\varrho}, \hat{\sigma}_{\mu}^{-} ]  + [ \sigma_{\mu}^{+}, \hat{\varrho}\hat{\sigma}_{\mu}^{-} ] )  + \tfrac{1}{4} \gamma_{p}  ( [ \hat{\sigma}_{\mu}^{z}, \hat{\varrho}\hat{\sigma}_{\mu}^{z} ] + [ \hat{\sigma}_{\mu}^{z}\hat{\varrho}, \hat{\sigma}_{\mu}^{z} ] ) \, , \\
      \hat{\mathcal{L}}_{\mathrm{FB}} [\hat{\varrho}] & = \tfrac{1}{2} c \kappa  \sum_{k}
     ( 1+ \bar{n}) ( [\hat{a}_{k}  \hat{\varrho}, \hat{a}_{k}^{\dag} ] + [ \hat{a}_{k}, \hat{\varrho}\hat{a}_{k}^{\dag} ]) + \bar{n}([ \hat{a}_{k}^{\dag}\hat{\varrho}, \hat{a}_{k}] + [ \hat{a}_{k}^{\dag}, \hat{\varrho}\hat{a}_{k} ] ). 
  \end{aligned}
\end{equation}
\end{widetext}
Here, $W_{21}$ is the relaxation rate from the excited to the ground state, $W_{12}$ is the incoherent pumping rate, and $\gamma_{p}=3\gamma_{0}$ is the pure dephasing rate. For the field, $\kappa$ is the absorption rate during the propagation within the medium and 
\begin{equation}
\label{photonnumb}
\bar{n}= \frac{1}{\exp\left (\frac{\hbar \omega_{0}}{k_{B}T_{f}}\right )-1}
\end{equation} 
is the mean equilibrium photon number in each reservoir mode of interest with $T_{f}$ to be temperature of the field background reservoir. Note that for a PCF,  $\bar{n}$ will be the closer to zero the lower the losses are. In that sense $\bar{n}$ is determined by the quality of the PCF.

If we consider the thermal temperature of the radiative reservoir for the atoms to be $T_{a}$, then:
\begin{align}
    W_{21}=\gamma_{0} (1+ \bar{n}_{a} )\, ,  \qquad  \qquad   W_{12} = \gamma_{0} \bar{n}_{a} \, ,
\end{align}
with photon occupation number $\bar{n}_{a}$ given by \eqref{photonnumb} with temperature $T_{a}$.  

It is customary to define the longitudinal and transverse damping rates as
\begin{equation}
   \gamma_{\|} = W_{12} +W_{21} \, , \qquad \qquad 
    \gamma_{\perp} = \gamma_{p}+\tfrac{1}{2} \gamma_{\|}  \, .  
\end{equation}
These coefficients $\gamma_{\|}$ and $\gamma_{\perp}$ correspond to two different damping mechanisms; namely, longitudinal (population decay) and transverse (dephasing).  In the case of population decay, the excited atoms have a spontaneous tendency to decay to the ground state. Since this is a stochastic process, it randomly breaks the coherence of the light field. Consequently, a spontaneous emission decay would constantly interrupt the Rabi oscillations. On other hand, transverse damping process causes the excited atoms to undergo an elastic or near-elastic collision, which breaks the phase of the light pulse without modifying the population of the excited state. Eventually, the effect of randomizing the phase of the light field, will destroy the Rabi oscillations.  

\subsection{Dynamics of the positive $P$ distribution}

We define an optical field $\hat \Omega(\mathbf{r})$ and collective operators for the atoms within the $j$th volume element and $m$th frequency band as follows:
\begin{equation}
\begin{aligned}
    \label{Eq:collective_opr}
    & \hat{\Omega}(\mathbf{r}_j)   = 2ig\sqrt{\frac{V}{\Delta V}} \sum_k \hat{a}_k e^{i k \cdot r_j}, \\ 
   & \hat{R}^{z}(\mathbf{r}_j, \omega_m) = \frac{1}{N_{jm}} \sum_{n}^{N_{jm}} \hat{\sigma}_{jmn}^{\sigma} ,  
\end{aligned}
\end{equation}
where the superscript $\sigma$ takes the values $z, \pm$ and $N_{jm}$ is the number of two-level atoms in the $j$th volume in frequency band centered at $\omega_m$. Thus, $\rho(\mathbf{r}_j,\omega_m)$ is now a combined frequency and spatial density in the small volume. The variables correspond to the Rabi frequency and Bloch vector components, respectively.  

Since direct numerical simulation of the master equation for a $N$ two level-atom system is extremely difficult, our strategy is to derive the suitable equations of motion in phase space. As heralded in the Introduction, we use the positive $P$ approach~\cite{Drummond:1980aa}, which is a normally ordered operator representation such that identifies the moments of $\hat{\varrho}$ with the corresponding $c$-number moments of a positive $P$ distribution. 

In this approach, we have a mapping $\hat{\Omega} \leftrightarrow \Omega$, $\hat{\Omega}^{\dag} \leftrightarrow \Omega^{\dag}$, $\hat{R}^{\pm} \leftrightarrow R^{\pm}$, $\hat{R}^{z} \leftrightarrow R^{z}$ and, following the standard procedures, the master equation can then be transformed into an equivalent Fokker-Planck equation for $P(\Omega, \Omega^{\dag}, R^{-}, R^{+}, R^{z})$. This equation is valid only when the distribution $P(\Omega, \Omega^{\dag}, R^{-}, R^{+}, R^{z})$ vanishes sufficiently rapidly at the boundaries. In practical applications, it is usually the case that the damping terms provide a strong bound at infinity on the distribution function~\cite{Gilchrist:1997aa}.

In terms of these variables, and in the limit of large $N$, we get the following set of the stochastic equations which serve as the basis for the simulation:
\begin{widetext}
\begin{equation}
 \begin{aligned}
   \label{field_R_6}
   \left ( \frac{\partial}{\partial  \mathbf{r}}+\frac{1}{c}\frac{\partial}{\partial t} \right ) \Omega(t, \mathbf{r}) & = -\frac{1}{2}\kappa \Omega(t, \mathbf{r}) + G \int\rho( \mathbf{r},\omega)R^{-}(t, \mathbf{r},\omega)d\omega +F^{\Omega}(t, \mathbf{r}), \\
   \frac{\partial}{\partial t}R^{-}(t, \mathbf{r},\omega) & = - (\gamma_{\perp} + i(\omega -\omega_{0}))R^{-}(t, \mathbf{r},\omega)  + \Omega(t, \mathbf{r})R^{z}(t, \mathbf{r},\omega)+F^{R}(t,  \mathbf{r}, \omega), \\
 \frac{\partial}{\partial t}R^{z}(t, \mathbf{r},\omega) & = - \gamma_{\|} [R^{z}(t, \mathbf{r},\omega)-\sigma^{SS} ] -\frac{1}{2} [ \Omega(t, \mathbf{r})R^{+}(t, \mathbf{r},\omega) +\Omega^{\dag}(t, \mathbf{r})R^{-}(t, \mathbf{r},\omega) ]+F^{z}(t,  \mathbf{r}, \omega),  
 \end{aligned}
 \end{equation}
 \end{widetext}
where 
\begin{equation}
\sigma^{SS} = \frac{W_{12}-W_{21}}{W_{12}+W_{21}} \,,
\qquad  
G=\frac{V g^{2}}{c} \, .
\end{equation}
Equations~\eqref{field_R_6} are identical with the usual semiclassical equations for the slowly varying envelope fields~\cite{Lax:1966aa,Haken:1966aa}, except for the presence of the Langevin terms $F$ that describe quantum fluctuations and depend on the bath and nonlinear atom-field coupling, and are expressed as:
\begin{widetext}
    \begin{equation}
\begin{aligned}
     F^{\Omega}(t, \mathbf{r}) & = 2\xi^{\alpha}(t, \mathbf{r}) \sqrt{G\kappa \overline{n}}= [F^{{\Omega}^{\dag}}(t, \mathbf{r})]^{\ast}, \\
      F^{R}(t,  \mathbf{r},\omega) & = \frac{1}{\sqrt{\rho(\mathbf{r}, \omega)}} \{ \xi^{J}(t, \mathbf{r}, \omega) \sqrt{2\Omega R^{-}} + 2\xi^{P}(t, \mathbf{r},\omega)\sqrt{\gamma_{P}(R^{z}+1)} +2\xi^{o}(t, \mathbf{r}, \omega)\sqrt{W_{12}}\}, \\
       F^{R^{\dag}}(t,  \mathbf{r}, \omega) & = \frac{1}{\sqrt{\rho(\mathbf{r}, \omega)}} \{ \xi^{J^{\dag}}(t,  \mathbf{r}, \omega) \sqrt{\Omega^{\dag}R^{+}} +2\xi^{P\ast}(t,  \mathbf{r}, \omega)\sqrt{\gamma_{P}(R^{z}+1)}    +2\xi^{o\ast}(t,  \mathbf{r}, \omega)\sqrt{W_{12}} \} , \\
        F^{z}(t,  \mathbf{r}, \omega) & =\frac{1}{\sqrt{\rho(\mathbf{r}, \omega)}} \{ \xi^{z}(t,  \mathbf{r}, \omega)[(2\gamma_{\parallel})(1 - \sigma^{SS}R^{z}) + (R^{-} \Omega^{\dag} - R^{+} \Omega)-2W_{12} R^{+}R^{-}]^{1/2}  - [\xi^{o}(t,  \mathbf{r},\omega) R^{+} + \xi^{o\ast}(t,  \mathbf{r},\omega) R^{-}] \sqrt{W_{12}} \} .
\end{aligned}
\end{equation}
\end{widetext}
The terms; optical thermal noise  $\xi^{\alpha}(t, \mathbf{r})$, incoherent pumping noise $\xi^{o}(t, \mathbf{r},\omega)$ and collisional dephasing noise $\xi^{P}(t, \mathbf{r}, \omega)$ are complex, while photon-atom interaction noise $\xi^{J}(t, \mathbf{r},\omega), \xi^{J^{\dag}}(t, \mathbf{r},\omega), \xi^{z}(t, \mathbf{r},\omega)$ are real.  The correlation properties are
\begin{widetext}
\begin{equation}
\begin{aligned}
    &\langle{\xi^{\alpha}(t, \mathbf{r}) \xi^{\alpha \ast}(t^{\prime},  \mathbf{r}^{\prime})}\rangle = \delta(t-t^{\prime}) \delta^{(3)}( \mathbf{r}- \mathbf{r}^{\prime}),\\
    &\langle{\xi^{o}(t, \mathbf{r},\omega) \xi^{o \ast}(t^{\prime}, \mathbf{r}^{\prime},\omega^{\prime})}\rangle = \delta(t-t^{\prime})  \delta^{(3)}( \mathbf{r}- \mathbf{r}^{\prime})\delta( \omega- \omega^{\prime}) , \\
    &\langle{\xi^{P}(t,  \mathbf{r}, \omega) \xi^{P \ast}(t^{\prime}, \mathbf{r}^{\prime},\omega^{\prime})}\rangle = \delta(t-t^{\prime})  \delta^{(3)}( \mathbf{r}- \mathbf{r}^{\prime})\delta( \omega- \omega^{\prime}),\\
    &\langle{\xi^{J}(t, \mathbf{r}, \omega) \xi^{J}(t^{\prime}, \mathbf{r}^{\prime}, \omega^{\prime})}\rangle = \delta(t-t^{\prime})  \delta^{(3)}( \mathbf{r}- \mathbf{r}^{\prime})\delta( \omega- \omega^{\prime}),\\
    &\langle{\xi^{J^{\dag}}(t,  \mathbf{r}, \omega) \xi^{J^{\dag}}(t^{\prime}, \mathbf{r}^{\prime},\omega^{\prime})}\rangle = \delta(t-t^{\prime})  \delta^{(3)}( \mathbf{r}- \mathbf{r}^{\prime})\delta( \omega- \omega^{\prime}), \\
    &\langle{\xi^{z}(t,  \mathbf{r}, \omega) \xi^{z}(t^{\prime},  \mathbf{r}^{\prime},\omega^{\prime})}\rangle = \delta(t-t^{\prime})  \delta^{(3)}( \mathbf{r}- \mathbf{r}^{\prime})\delta( \omega- \omega^{\prime}).
\end{aligned}
\end{equation}
\end{widetext}
 Since the equations are derived through a normally ordered representation, there are bath noise terms associated with dephasing ($\gamma_p$) and gain ($W_{12}$), but not loss ($W_{12}$).  Furthermore, the gain noise is only present at finite temperatures.  In addition to the bath noise, the positive-$P$ method has noise associated with the atom-field coupling, which is present even for unitary evolution and corresponds in some sense to shot-noise effects in the atom-light interaction.

\section{Results}
\label{sec:res}

\subsection{Amplitude Squeezing}

To solve the system~\eqref{field_R_6} it proves convenient to use a propagating reference frame 
moving with the center of the pulse in propagation direction $z$ with the velocity $v_{g}$, and thus involving a retarded time~\cite{kozlov:2000aa,corney:2008aa}~$\tau=t-z/v_{g}$. In this retarded frame, the first equation in~\eqref{field_R_6} becomes:
\begin{align}
\left [ \frac{\partial}{\partial z}  + \left (\frac{1}{c}-\frac{1}{v_{g}} \right ) 
\frac{\partial}{\partial \tau} \right ] \Omega(\tau,z)  =  -\frac{1}{2} \kappa \Omega(\tau,z) \nonumber  \\
+ G \int\rho(z,\omega)R^{-}(\tau,z,\omega)d\omega + F^{\Omega}(\tau,z).
\end{align}
For a coherent field the appropriate initial condition is
\begin{align}
    P (\tau, \Omega, \Omega^{\dag})= 
    \delta^{(2)}(\Omega^{\ast} -\Omega^{\dag}) 
    \delta^{(2)} [ \Omega - \mathcal{E}(0,\tau) ],
    \end{align}
with $\mathcal{E}(z,\tau)$ being the soliton shaped pulse in the retarded time frame, 
\begin{equation}
\begin{aligned}
    \mathcal{E}(z, \tau) & = 2 A \cosh^{-1}\left[A(\tau-\tau_{0})  \right]\exp\left[i(\delta\tau+\phi(z)) \right],   \\
    \phi(z) & = \frac{\delta}{A^2 + \delta^{2}}G\rho z \, ,
\end{aligned}
\end{equation}
where $2A$ is the pulse amplitude, $\tau$ the pulse timing, $\delta$ is detuning, $\phi(z)$ is the phase and $v_{g}$ obeys:
\begin{align}
    \frac{1}{v_{g}}=\frac{1}{c} + \frac{1}{2} \frac{G\rho}{A^2+\delta^{2}} \, .
\end{align}
For the atoms initially distributed in the ground state we have the initial condition
\begin{equation}
R^{+} = R^{-}=0 , \qquad  R^{z}=-1/2 .
\end{equation}

\begin{figure*}[t]
    \includegraphics[width=1.85\columnwidth]{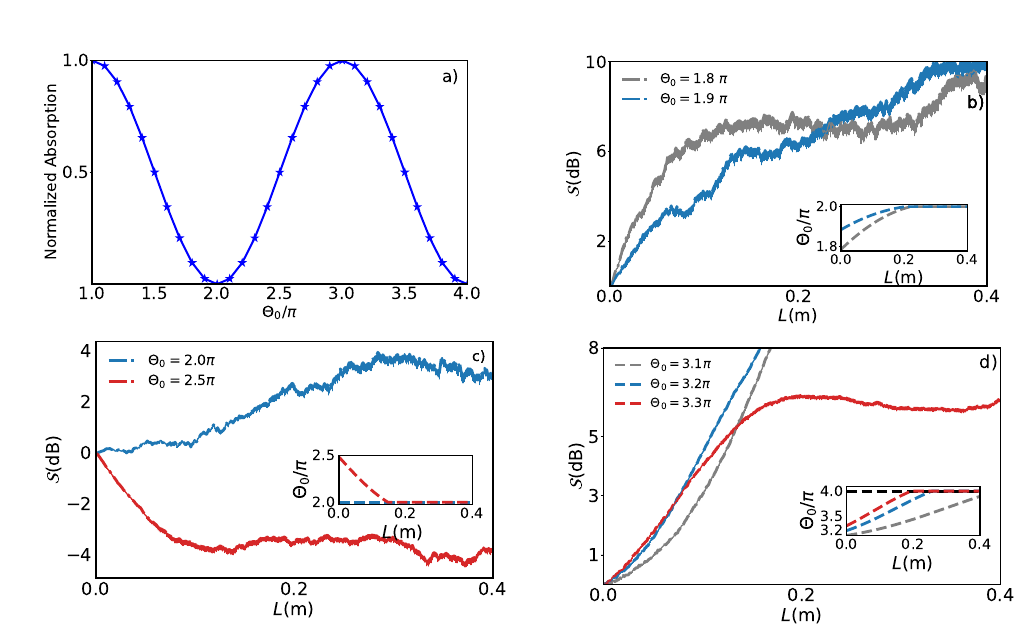}
    \caption{\label{fig:Fig2} a) The normalized absorption of atoms (blue solid line) versus area. When $\Theta$ takes values of $\pi$ and 3$\pi$, indicating maximum energy transfer from the pulse to atoms. However, at $\Theta = 2\pi$ and $\Theta = 4\pi$, the absorption drops to zero, signifying no energy transfer. In the range between $\pi$ and 2$\pi$, as well as between 3$\pi$ and 4$\pi$, the absorption of atoms decreases to zero, allowing pulse energy to increase. When $\Theta$ ranges from 2$\pi$ to 3$\pi$, the absorption of atoms rises back to 1, indicating energy transfer to atoms.
    b) Excessive noise as a function of the fiber length for $\Theta=1.8\pi, 1.9\pi, 2.0\pi$ (the inset subplot indicates the growth of the pulse area during propagation along the fiber). c) Comparison of the squeezing of an input pulse with an initial area $2.5\pi$ during propagation into the fiber, in the presence and absence of longitudinal damping. The transparent shaded area around the dashed blue/red line indicates the uncertainty of the squeezing value in each grid point.  d)  Squeezing for pulses with $\Theta=3.1\pi, 3.2\pi, 3.3\pi$ (the subplot shows the pulse area propagation within the fiber before and after it reaches to the stable area $\Theta=4\pi$). The atom number in each cell is $N=1000$ and temperature is considered to be zero in the subplots a to c.}
\end{figure*} 

We next concentrate at amplitude squeezing, which is more easily detectable. To calculate amplitude squeezing, it is necessary to define the normalized variances based on the energy of the output field at the point z along the fiber
\begin{align}
    \hat{M}(z)=\int_{-\infty}^{\infty}  \hat{\Omega}^{\dag}(\tau,z)  \hat{\Omega}(\tau,z) \,  d\tau ,
\end{align}
with the corresponding squeezing ratio
\begin{align}
    S(z)= \frac{\mathrm{Var} [\hat{M}(z)]}{\mathrm{Var}  [\hat{M}(0) ]} \, ,
\end{align}
where $\mathrm{Var} [\hat{M}(z) ]= {\langle \hat{M}^2(z)\rangle} - {\langle \hat{M}(z) \rangle}^2$.

To calculate the variance with a $+{P}$ simulation, we need to express it in terms of normally ordered correlations:
\begin{align}
    \hat{M}^{2}\simeq \, :\hat{M}^{2}: + \frac{4g^2 L}{v_g} \hat M \, ,
\end{align}
 where we have used the approximate equal-space commutation relation $ [ \hat{\Omega}(\tau,z) , \hat{\Omega}^{\dag}(\tau', z)]\simeq 4g^{2}L\delta(\tau-\tau^{'})/v_g$. The squeezing ratio is then
\begin{align}
    S = 1+ \frac{v_g\mathrm{Var}_{+P}
   [\hat{M}]}{4 g^{2}L \langle \hat M\rangle}\, ,
\end{align}
where $\mathrm{Var}_{+P} \equiv {~:\langle \hat{M}^{2} \rangle:} - {\langle \hat{M} \rangle}^{2}$.

The area of a pulse is defined as~\cite{Haus:1989aa} 
\begin{equation}
    \Theta(z) = \int \Omega(\tau, z) d\tau \, .
\end{equation} 
For a hyperbolic secant soliton $\Theta=2\pi$, the pulse shape remains unchanged during propagation. However, any initial pulse area $\Theta_{0}$ that satisfies $(m+1)\pi>\Theta_0 m\pi$, will grow in area towards $(m+1)\pi$ if $m$ is an odd number or it will shrink in area towards $m\pi$ if $m$ is an even number.  Figure~\ref{fig:Fig2}a) shows the total absorbed energy by atoms as a function of area.

Since when $\Theta_0 =2\pi$ the pulse amplitude remains unchanged, there is no reduction in the fluctuations of the solitons's amplitude. Consequently, no amplitude squeezing  occurs.  For pulses $\pi < \Theta_0 < 2\pi$, the  area will grow until it reaches the stable value $\Theta=2\pi$. This leads to an increase in the amplitude fluctuations leading to no amplitude squeezing; as can be appreciated in Figure.~\ref{fig:Fig2}b).

However, if we consider a pulse with $\Theta_0 = 2.5\pi$, pulse area is not stable during the propagation and it diminishes towards $\Theta = 2\pi$. As a result, the pulse loses energy and leaves the atoms excited. Our simulation shows that this reduction in the pulse area results in diminished amplitude's fluctuations, ultimately leading to amplitude squeezing.

Figure~\ref{fig:Fig2}c) displays the calculated amplitude squeezing as a function of fiber length for two different initial pulse areas. The results reveal a precisely 2$\pi$ soliton pulse exhibits no amplitude squeezing during the pulse propagation along the fiber. Whereas a pulse with an initial area of $2.5\pi$ undergoes squeezing after propagating a distance of $z=0.4$~m, but before reaching to the stable area $2\pi$. The simulation reveals amplitude squeezing of approximately -4~$\mathrm{dB}$ in this case. It is important to note that in Fig.~\ref{fig:Fig2}c), temperature and both longitudinal and transverse damping are assumed to be zero. 

In contrast, when the initial area is considered $\Theta_0 >3\pi$, the pulse area intensifies until it reaches to $4\pi$. Hence the pulse amplitude fluctuations increase, and no squeezing is observed; see Fig.~\ref{fig:Fig2}d).

We can think of the results as the consequence of pulse reshaping area. When pulse area shrinks toward $m\pi$ where $m$ represents an even number, the fluctuations amplitude  compress and undergo squeezing. On the other hand, when the pulse expands towards $m\pi$, the fluctuations in the amplitude encounter anti-squeezing. 

\begin{figure}[t]
    \centering
    \includegraphics[width=.90\columnwidth]{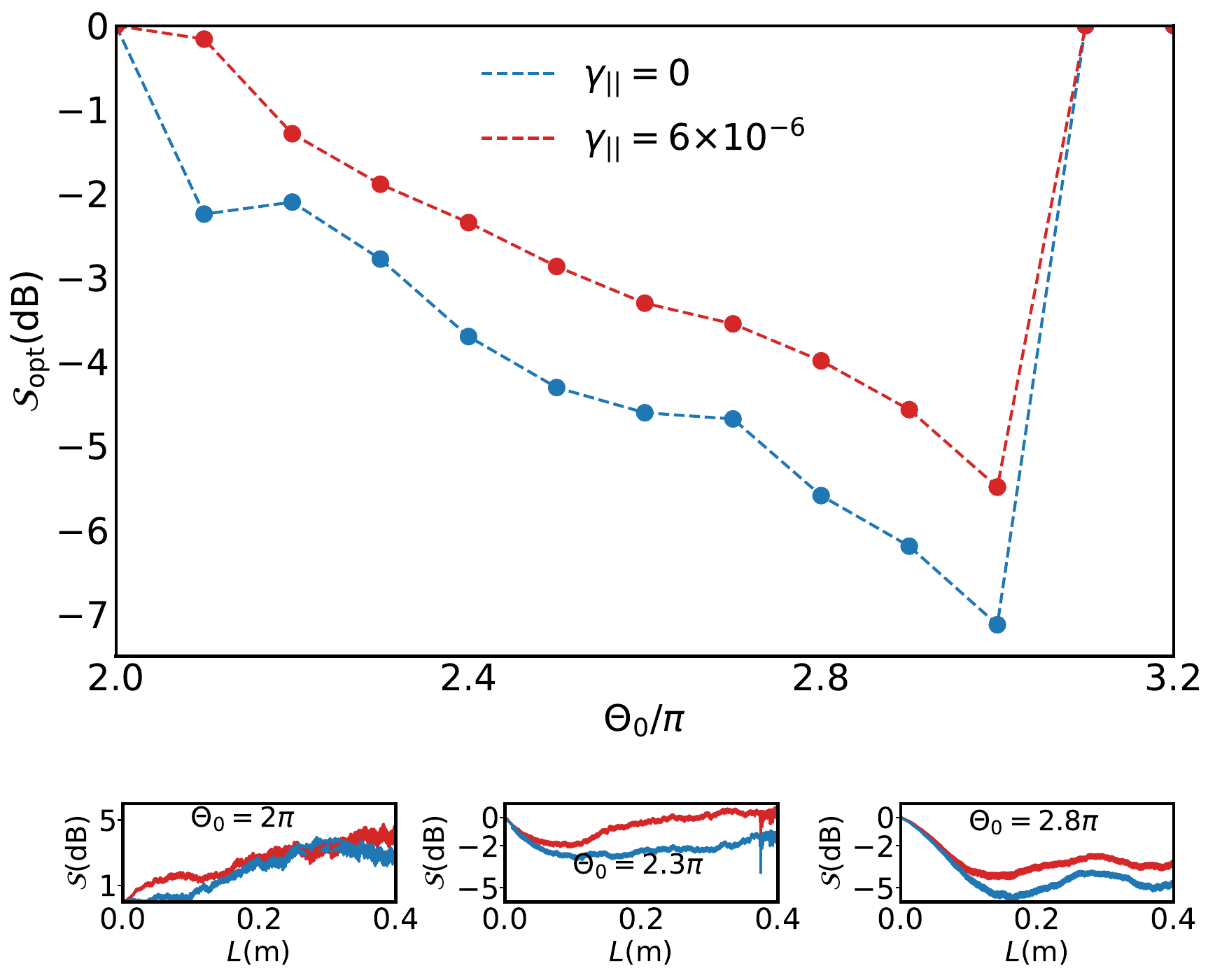}
    \caption{ \label{fig:Fig3} Optimum squeezing as a function of the initial area of the input pulse in the presence and absence of longitudinal damping. The bottom subplots indicate the evolution of squeezing for pulses with the initial area $2\pi$, $2.3\pi$, and $2.8\pi$. The blue curves are in the absence of the longitudinal damping while the red curves capture the effect of damping. The transparent shaded color shows the uncertainty of the achieved squeezing from 4000 samples in each grid point. The longitudinal damping rate is taken to be $\gamma_{\|}=1 \times 10^{-6}$, while temperature is kept to zero.  }
\end{figure}

\subsection{Optimal squeezing}
We next examine the effect of the longitudinal damping $\gamma_{\|}$ on the amplitude squeezing. Specifically, we compare the results when the longitudinal damping is treated as $\gamma_{\|}=\gamma_{0}$ with the previous scenario where longitudinal damping was assumed to be zero.

\begin{figure}[t]
    \includegraphics[width=0.90\columnwidth]{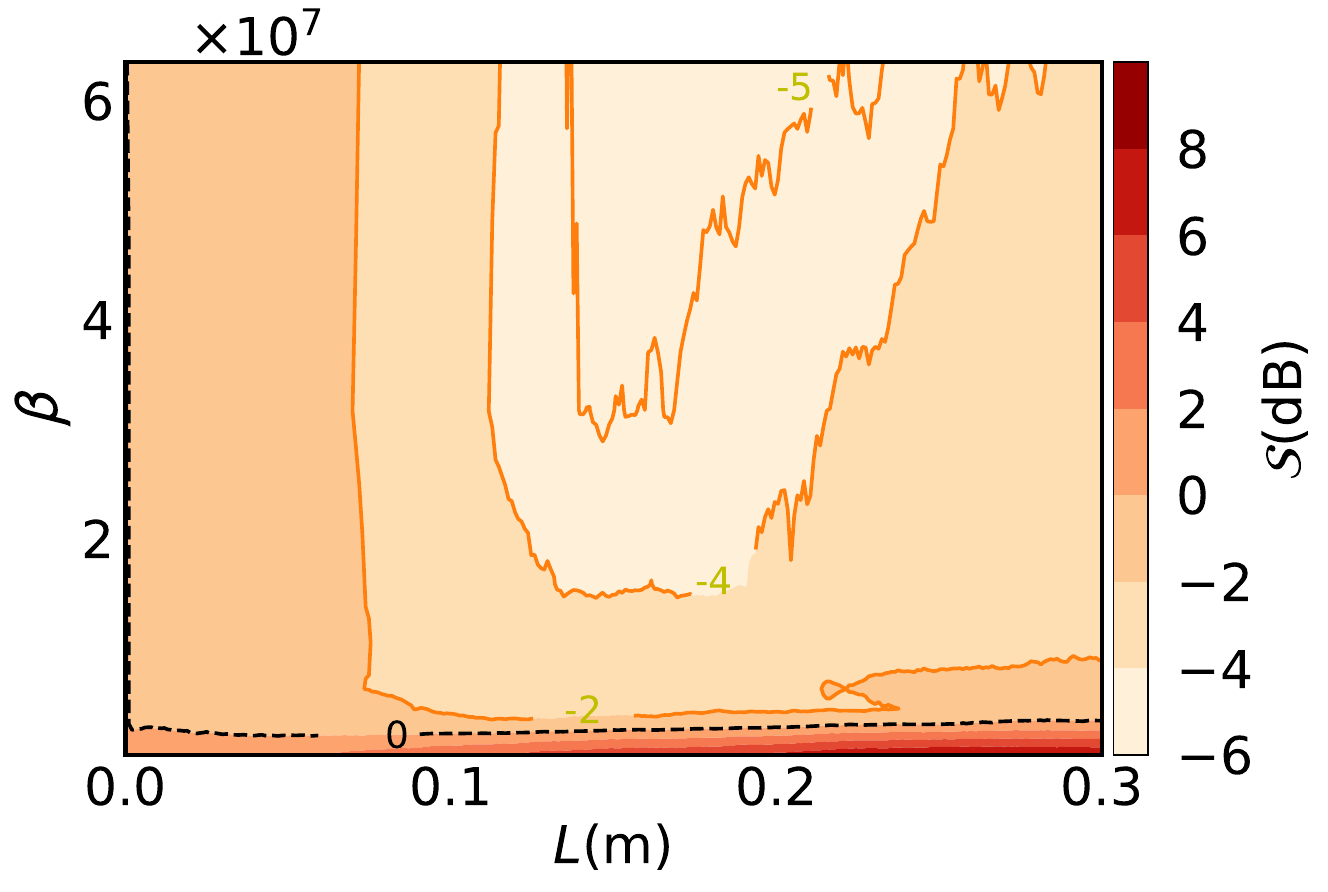}
    \caption{ \label{fig:Fig4}The impact of temperature on the squeezing is shown. The squeezing is calculated over 2000 samples in this dataset.}
\end{figure}
Figure~\ref{fig:Fig3} illustrates the optimal squeezing as a function of the pulse area. The optimal squeezing in the presence and absence of longitudinal damping is calculated. The error bar on each data point indicates the optimal squeezing achieved from 4000 samples. It is evident that taking into account $\gamma_{\|}$ leads to a decrease in the amplitude squeezing. Longitudinal damping enhances the decay of atoms into the ground state and this continuous decay process disrupts the Rabi frequency and contributes to increased amplitude pulse fluctuations, resulting in the reduction of amplitude squeezing observed in the system.  Since temperature is taken as zero in Fig.~\ref{fig:Fig3}, there is no contribution of thermal noise. The remaining sources of noise originate from the dipole-field interaction and spontaneous emission.
\begin{figure}[b]
    \includegraphics[width=.90\columnwidth]{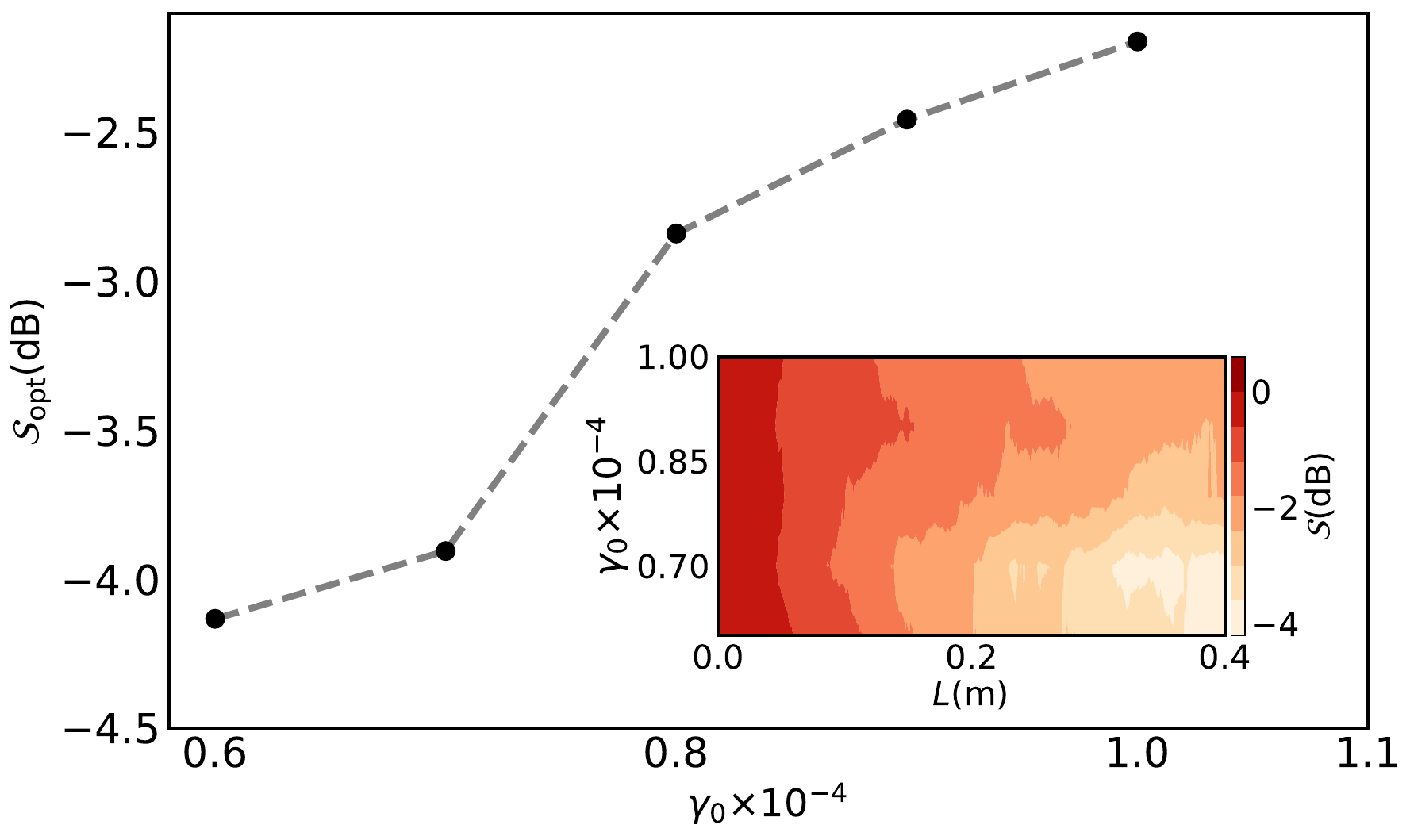}
    \caption{ \label{fig:Fig5}
    The impact of spontaneous emission rate $\gamma_{0}$ on the amplitude squeezing is shown.
    The plot displays the optimum value of the amplitude squeezing ($S_{\mathrm{opt}}$) achieved from 2000 samples for a range of $\gamma_{0}$ between $0.6\times10^{-4}~1.1\times10^{-4}$. The subplot indicates the evolution of squeezing within the fiber for each $\gamma_0$. The initial area of the input pulse is set to be $2.3\pi$. The thermal noise is kept zero.}
\end{figure}
Looking at Fig.~\ref{fig:Fig2}a), the pulse energy loss has its maximum at $\Theta=3\pi$, the incident pulse with $\Theta_0=3\pi$ undergoes the greatest reshaping before it reaches to the stable area; $\Theta_0=2\pi$. Consequently, starting with a pulse with $\Theta_0 =3\pi$, results in the most pronounced reduction of amplitude fluctuations, and eventually leading to the squeezing dip.

Furthermore, the effect of damping on the optimum squeezing for each pulse is shown in the subplots in Fig.~\ref{fig:Fig3}. The three subplots indicate the squeezing in the presence and absence of longitudinal damping within the fiber for $\Theta_0=2\pi$, $\Theta_0=2.3\pi$ and $\Theta_0=2.8\pi$. Comparing the two pulses with $\Theta_0=2.3\pi$ and the other  with an area of $2.8\pi$, the pulse with $2.3\pi$ reaches the stable area sooner. As a result, the optimum squeezing ($\mathcal{S}_{\mathrm{opt}}$) occurs over a shorter distance for the $2.3\pi$ pulse, as illustrated in Fig.~\ref{fig:Fig3}.

In Figure~\ref{fig:Fig4}, the impact of temperature on the squeezing is demonstrated for a pulse with $\Theta_0=2.8$. In this case all sources of noise, including thermal noise and spontaneous emissions are taken into account. In addition, atoms experience a Doppler broadening as the temperature increases.  In terms of  $\beta = \hbar \omega_{0}/K_{B}T$, where $\omega_{0} = 30 GHz$, it is observed that the highest level of squeezing occurs within the range $\beta = 2-6\times 10^{7}$ which corresponds to temperatures in the range of nano Kelvins. Considering this, it's essential to note that the dataset is not optimized. Depending on the system parameters, such as atoms and pulse characteristics, one might achieve a different optimal regimes for the amplitude squeezing.

The effect of small spontaneous emission rate on the amplitude squeezing exhibits a resemblance to the impact of damping. Figure~\ref{fig:Fig5} illustrates the optimal squeezing achieved for various strengths of $\gamma_{0}$. In the weak $\gamma_{0}$ regime, a squeezing level of -4 dB is attained. The subplot in Fig.~\ref{fig:Fig5} illustrates how squeezing evolves concerning $\gamma_{0}$ along the fiber. Within the weak $\gamma_{0}$ strength regime, the highest amplitude squeezing occurs towards the fiber's end. There are two primary reasons for this. Firstly, weak $\gamma_{0}$ results in a smaller coupling strength, causing the pulse to move more slowly through the medium. Secondly, this condition also reduces the medium's noise level. However, in the moderate regimes, the squeezing significantly diminishes.

\section{Conclusions}
\label{sec:conc}

Resorting to the formalism of the positive $P$ function, we have developed  a full quantum model of SIT that has allowed us to characterize the dynamic behavior of amplitude squeezing. Our investigation encompassed both the influence of the thermal and quantum noise. Our results demonstrate a strong dependence of amplitude squeezing in SIT solitons on both the initial pulse area and the absorbed energy. Furthermore, we have shown that damping can noticeably diminish squeezing, even when temperature effects are negligible. However the primary detrimental impact arises from the thermal noise, leading to a complete suppression of squeezing.

\acknowledgments
We are indebted to Ray Kuang Lee and Thomas Dirmeier for discussions. L.L.S.S. acknowledges financial support from  Spanish MINECO (Grant No. PID2021-127781NB-I00).  A.A.S. acknowledges financial support from the Ministry of Science
and Higher Education of the Russian Federation (Grant No. 075-15-2022-316) and from the Foundation for the Advancement of Theoretical Physics and Mathematics “BASIS”.

\bibliography{SIT}

\end{document}